\documentclass[epj]{svjour}
\usepackage{graphics}

\begin{document}
\title{Shake induced order in nanosphere systems}
\author{F. J\'{a}rai-Szab\'{o}\inst{1} \and Z. N\'{e}da\inst{1} \and
S. A\c{s}tilean\inst{1} \and C. Farc\u{a}u\inst{1} \and A.
Kuttesch\inst{1}}
%
%
\institute{Babe\c{s}-Bolyai University, Department of Physics, Str.
Kog\u{a}lniceanu 1, RO-400084, Cluj-Napoca, Romania}
\date{Received: date / Revised version: date}
%
\abstract{ Self-assembled patterns obtained from a drying nanosphere
suspension are investigated by computer simulations and simple experiments.
Motivated by the earlier experimental results of Sasaki \& Hane and Sch\"ope, we confirm that more
ordered triangular lattice structures can be obtained whenever a moderate intensity 
random shaking is applied on the drying system. Computer simulations are realized on an improved
version of a recently elaborated Burridge-Knopoff type model. Experiments are made following the
setup of Sasaki and Hane, using ultrasonic radiation as source for controlled shaking. 
} 
\maketitle
\section{Introduction}
\label{intro}

In order to use less material, to obtain a more effective storage of
information or easier access to tight and narrow spaces, modern
areas of engineering needs structures or devices as small as
possible. Nanostructures became ideal candidates in this sense.
Thanks to the efforts of nano-chemists, nowadays various
nanoparticles nearly monodisperse in terms of size, shape, internal
structure, and surface chemistry, can be produced through reliable
and standard manufacturing processes. These nanoparticles can be
nanospheres, nano-tubes or colloids, and can be used as building
blocks for engineering more complex structures.  Human assisted
construction is however very complicated and unproductive. Designing
thus technologies where the building blocks self-assemble is of
primary importance. In the present contribution a self-assembling
system which is widely used in NanoSphere Lithography (NSL)
\cite{Denkov1992,Denkov1993,Haynes2001,Kempa2003,Murray2004,Chabanov2004,Vasco2004,Chen2005} is
studied. Inspired by the earlier experimental work of Sasaki \& Hane \cite{sasaki} and
Sch\"ope \cite{schope},
here it is confirmed by computer simulations and independent experiments 
that an extra shaking mechanism imposed on the drying system 
is benefic for engineering structures with better practical properties.

The studied system is a monodisperse polystyrene na\-no\-sphe\-re
suspension which is dried on a previously prepared silica substrate.
Self-assembly is governed by several forces: (i) capillarity forces,
(ii) electrostatic repulsion forces between the slightly negatively charged 
nanospheres (iii) hard-core type repulsion forces between the nanospheres and 
(iv) pinning forces acting on the nanospheres. Many
previous studies have shown that after the drying process is
completed the nanospheres self-organize in complex patterns. The
case when monolayer nanosphere patterns are obtained is of special
practical importance. These monolayer structures are useful as
deposition masks for NanoSphere Lithography (NSL).
NSL is nowadays recognized as a powerful fabrication technique to
inexpensively produce nanoparticle arrays with controlled shape,
size, and interparticle spacing \cite{Murray2004,Baia2006}.

\begin{figure}
\begin{center}
 \resizebox{8cm}{!}{
  \includegraphics{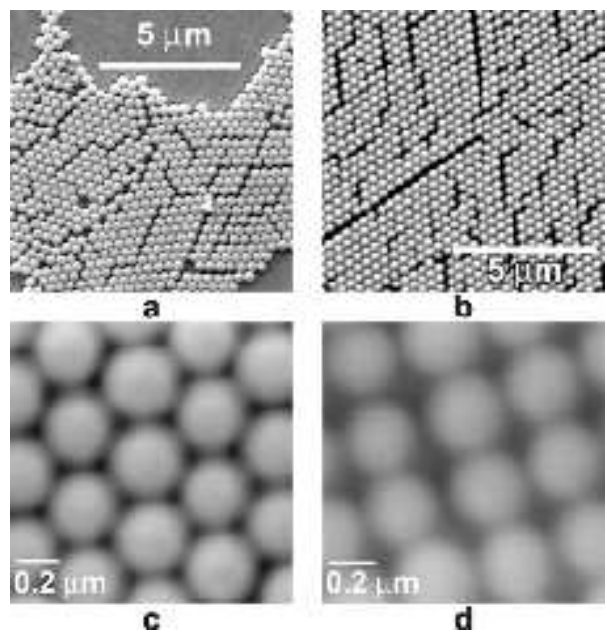}}
\end{center}
\caption{Characteristic self-organized patterns obtained after the
drying process is ended. The figures from the bottom line
illustrates the triangular and square crystallization phases. Results
from our own experiments, for experimental details see \cite{Baia2006}.}
\label{fig:1}
\end{figure}

Ideally, it is expected that the nanospheres will order in the
compact triangular lattice structure (Figure \ref{fig:1}c). However, apart of this ideal
lattice topology (desirable for most of the practical purposes),
many other structures are formed. Usually the patterns presents many
dislocations, voids or clusters (Fig. \ref{fig:1}a,b). Sometimes
instead of the triangular symmetry, the less compact square topology
is selected (Fig. \ref{fig:1}d).  A fundamental goal for further
progress in NSL is the development of experimental protocols to
control the interactions, and thereby the ordering of nanoparticles
on the solid substrates \cite{Chabanov2004,Jarai2005}. 
 Recently \cite{Jarai2005}, we
proposed a Burridge-Knopoff type model
\cite{Burridge1967} for understanding the pattern formation
mechanism in this system. The model proved to be successful in
reproducing the observed monolayer patterns, and clarified the
influence of some experimentally controllable parameters.  By 
further improving this model it is straightforward to show, 
that applying an extra
random force on the system will result in more ordered triangular
nanosphere structures. One can explain thus the interesting experimental 
results of Sasaki \& Hane \cite{sasaki} and Sch\"ope \cite{schope}, and get further evidences that
our Burridge-Knopoff model works fine for describing this self-assembling phenomena.
 
The present
paper will investigate thus by large-scale computer simulations this phenomena and
present also further experimental evidences in support of Sasaki \& Hane's and Sch\"opfe's results.
First our Burridge-Knopoff
type model \cite{Jarai2005} is presented and improved. Then simulations with an
extra random or periodic shaking force are performed. Finally, as an 
experimental exercise the random shaking is
realized by an ultrasound bath in contact with the
drying nanosphere system. Both the simulations and experiments
clearly confirms that the extra shaking is useful for engineering more
ordered structures.

\section{Computer simulations of the patterns}
\label{sec:2}

\begin{figure}
\begin{center}
 \resizebox{8.5cm}{!}{
  \includegraphics{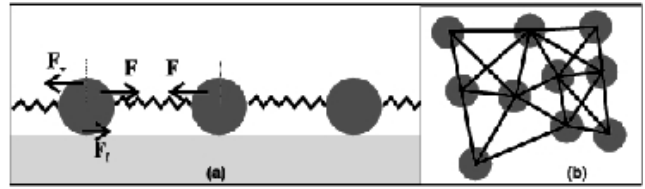}}
\end{center}
\caption{Basic elements of the spring-block stick-slip model.}
\label{fig:2}
\end{figure}

Pattern formation in the drying nano\-sphere suspension was recently
\cite{Jarai2005} successfully modeled by a 
relatively simple
Burridge-Kno\-poff type model \cite{Burridge1967,Cartwright1997}. The model is 
rather similar with the 
spring-block stick-slip model used for
describing fragmentation structures obtained in drying granular
materials in contact 
with a frictional substrate
\cite{Leung2000,Neda2002}. The new feature of the model
 presented in \cite{Jarai2005} is that a predefined lattice structure
is not imposed anymore. The model is two-dimensional; its main
elements are disks which can move on a frictional substrate and
springs connecting them (Figure \ref{fig:2}). Disks, all with the
same radius \textit{R}, model the nanospheres. The elastic
springs reproduces the resultant of all the forces acting between the nanospheres for 
small separation. These forces are the lateral capillarity forces  
and the electrostatic repulsion between the slightly negatively charged nanospheres. 
Springs have similar elastic constants \textit{k}, and their length
is defined as the distance between the perimeters of the connected
disks. There is an additional almost hard-core type repulsion $F_j$ which forbids disks to
interpenetrate. The friction (pinning) between disks and
surface equilibrates a net force less than $F_f$. Whenever the total force acting 
on a disk exceeds $F_f$, the disk slips with an over-damped motion. The tension in
each spring is proportional with the length of the spring $F_k =
kl$, and has a breaking threshold $F_b$.

It worth mentioning here that the approach based on the modified Burridge-Knopoff model seems realistic
only for dense nanosphere systems where the average separation between the nanospheres is less
than their radius. In this limit one can find a plausible argument for modeling the resultant forces between the nanospheres with harmonic forces. Before proceeding further with the description of the model we briefly present this argumentation.

\begin{figure}
\begin{center}
 \resizebox{8.5cm}{!}{
  \includegraphics{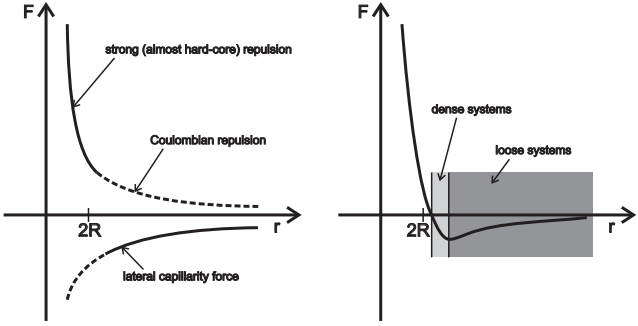}}
\end{center}
\caption{Forces acting between the nanospheres. On the image from the right-side the
resultant force is plotted. Positive force represents repulsion and negative forces  
attraction.}
\label{fig:3}
\end{figure}

Lateral capillarity forces have been studied in much detail both for micrometer size 
and nanometer size particles. 
Theoretical results \cite{chan1981,kralchevsky1993,paunov1993} 
predict that for relatively large 
separation between two spheres partially immersed in a liquid one would expect
an attractive force which is monotonically decaying with separation following a $1/r$
trend ($r$ the distance between the centers of the two spheres). Experimental 
results confirm this theoretical prediction for micrometer-sized particles \cite{dushkin1995,dushkin1996a,dushkin1996b}. For nanometer sized particles (our case) experiments were made only for the case when the two spheres are confined in a liquid film \cite{danov2001}. In this case the measured force 
had a non-monotonic variation with the separation of the particles. For small
distances it increases almost linearly, than reaches a maximum after which decays roughly as
$1/r$ with interparticle separation. In our case the polystyrene nanonspheres are slightly negatively charged, and there is thus an additional Coulombian repulsion between them. Assuming that the lateral capillarity force decay as $1/r$ and the Coulombian repulsion decay as $1/r^2$ one immediately gets that the resultant force has a non-monotonic variation and for small interparticle separation the resultant force should increases with $r$ (Figure \ref{fig:3}). The easiest way to model this resultant force for small interparticle separation is by assuming a linear variation with $r$. This trend is also in agreement with the lateral capillarity forces measured between nanospheres confined in a liquid film \cite{danov2001}. The above heuristic argument is far from being complete and lacks the numerical prefactors for the strength of the Coulombian repulsion and capillarity attraction forces. It serves only as a first motivation for the application of the simple spring-block approach.  In case of loose systems (where the interparticle separation is larger than the diameter of the nanospheres)  one should consider also forces decaying as $1/r$ and the relevance of the spring-block approach is questionable. It is interesting however, that even in this limit (see the pictures in \cite{Jarai2005}) the simple approach based on the Burridge-Knopoff type model works reasonably well.  In all simulations considered in the present work we will restrict our self on dense systems with interparticle separation much 
smaller than the particles diameter. In this case the harmonic force approach (modeling the resultant force with springs) is thus heuristically motivated and leads to realistic structures.   
   
In the simulation initially disks are randomly distributed and connected by a network
of springs with elastic constants $k_{ini}$. (Figure \ref{fig:2}b). We put springs between pairs of
disks, for which the centers can be connected without intersecting
another sphere (this condition will be referred later as the
geometric condition). An initially pre-stressed spring-block network
is thus constructed. During each simulation step the spring constant
is fixed and the system relaxes to an equilibrium configuration
where the tension in each existing spring is lower than the breaking
threshold $F_b$, and the total net force acting on each disk is
lower in magnitude than the slipping threshold $F_f$. This
relaxation is realized through several steps:
\begin{enumerate}
\item For all springs the tension $\left| \vec{F}^{ij}_k \right| $
is compared with $F_b$. If $\left| \vec{F}^{ij}_k \right| > F_b$,
the spring is broken and taken away from the system.

\item The total forces $\vec{F}^i_t =
\sum_p \left( d_{ip} \vec{F}^{ip}_k + \vec{F}^{ip}_j\right)$
acting on disks are calculated
(the sum is over all the other disks \textit{p}, $d_{ip}$ is 1 if
the disks are connected by a spring and 0 otherwise, the subscripts
\textit{k} and \textit{j} denotes elastic forces from springs and
hard-core type repulsion forces between disks, respectively).

\item Each disk is analyzed. If the magnitude of the total force
$\left|\vec{F}^i_t \right|$ acting on a disk is bigger than $F_f$,
then the disk will slip with an over-damped motion governed by
viscosity $\eta$, and its position will be changed by $d\vec{r} =
\vec{F}^i_t\,dt/\eta$. The repulsive hard-core
potential forbids the spheres to slide on each other and the
presence of viscosity eliminates unrealistic oscillations.

\item During the motion of a disk it can happen that another
spring is intersected. This intersected spring will brake and will
be taken away from the system.

\item After all disks have been visited in a random order and
their possible motions done, the springs that fulfill the considered
geometrical condition and for which the tension is lower than the
breaking threshold are redone. By this effect the rearrangement of
water between nanospheres is modeled.
\end{enumerate}

This concludes one relaxation step. The relaxation is continued
until a relaxation step is finished without having any spring
breaking or disk slipping event. After the relaxation is done, we
proceed to the next simulation step and increase all spring
constants by an amount \textit{dk}. This step models the phenomenon
that water level of the continuous film decreases due to evaporation
and the meniscus accounting for the capillarity forces gets more
accentuated. The system is relaxed for the new spring-constant
value, and the spring constant is increased again, until all springs
are broken or a stable limiting configuration is reached.

Several type of boundary conditions can be imposed. One possibility
is the free boundary condition which can be realized in a simple
manner by positioning initially the disks inside a circle to
minimize the effect of edges. Another possibility is to consider
fixed boundary conditions. This can be realized for example by
positioning again the disks inside a circle and considering a chain
of fixed disks on the chosen circle. These fixed disks are then
connected with the geometrically allowed springs to other disks. One
can also consider periodic boundary condition and position initially
the disks inside a rectangle.

The above sequence of events can be implemented on computers and
relatively big systems with over 10000 of disks can be simulated in
reasonable computational time. The model, as described above, has
several parameters: the value of the static friction force $F_f$,
the value of the breaking threshold $F_b$ of springs, the initial
value of spring constants $k_{\mathit{ini}}$, the spring constant
increasing step \textit{dk}, the value of viscosity $\eta$, the
parameters of the Lenard-Jones potential which realizes the hard-core 
repulsion, the radius of disks
\textit{R}, and the initial density of nanospheres $\rho = S/(N\pi
R^{2} )$ (where \textit{S} is the simulation area and \textit{N} is 
the number of considered disks). Varying these
parameters in reasonable limits all experimentally obtained patterns
can be successfully modeled \cite{Jarai2005}. 

An immediate question that arises in connection with the approach based
on the spring-block model concerns it's computational efficiency. What do we gain
by using this simplified force pattern instead of the more complete
variation sketched on Figure \ref{fig:3}? The answer comes from several 
sights. First, the exact shape of the force given in Figure \ref{fig:3} is 
undetermined since the two prefactors for the Coulombian and
capillarity forces are unknown. Due to this, there would appear an extra unknown parameter which 
determines the shape of the resultant force, making the model more complex. Secondly, the breaking 
threshold imposed on the springs allows the use of a natural cutoff for the interparticle forces, 
and makes simulations faster. Third, the simple harmonic force allows a more rapid calculation 
of the resultant forces. Finally, the geometric condition imposed on springs (not to intersect another nanosphere) insures the screening effect imposed by other nanospheres on capillarity 
and Coulombian forces.

\begin{figure}
\begin{center}
 \resizebox{8.5cm}{!}{
  \includegraphics{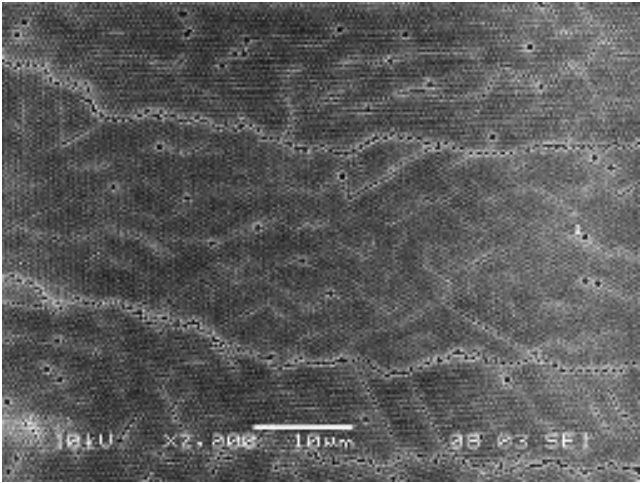}}
\end{center}
\caption{Characteristic dried nanosphere pattern obtained experimentally after
the drying front has moved in the horizontal direction.}
\label{fig:4}
\end{figure}

The above mentioned model considered a uniform decrease of the
liquid level in the whole system. Experimental studies show however
that this approach is not valid in real systems. During the drying
process large liquid level gradients are present, leading to the
flow of the nanosphere suspension from one part of the sample to
another one \cite{Denkov1992,Denkov1993}. In most of the experimental 
samples thus drying fronts
can be observed separating already dried and wet parts of the
sample. This non-uniform liquid level is also facilitated by almost
all type of suspension deposition methods on the silica substrate.
The obtained larger-scale patterns are strongly influenced by the
drying front. Experiments have shown that the characteristic larger-scale 
fracture-lines tend to be perpendicular on the direction of the drying front 
(Figure \ref{fig:4}).

Recently a simple lattice-gas type model was proposed in order to explain 
the self-assembly process driven by the drying liquid front \cite{Rabani2003}. 
Monte Carlo simulations with the classical Metropolis algorithm  
proved that patterns resembling the experimentally obtained ones can be obtained 
by using this elegant approach.

\begin{figure}
\begin{center}
 \resizebox{8.5cm}{!}{
  \includegraphics{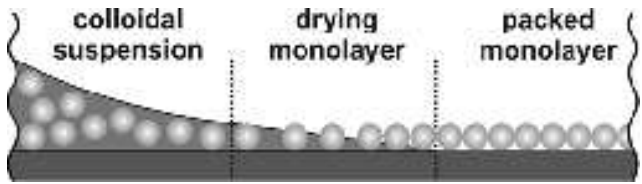}}
\end{center}
\caption{The three different regions in the improved model.}
\label{fig:5}
\end{figure}

Our model can be also further improved and made thus more realistic to
incorporate the effects generated by the moving drying front. In
order to do this we consider three different regions inside the
simulated area (Figure \ref{fig:5}).

In region I. the liquid level is greater than the na\-no\-sphe\-re
diameter, no capillary effects are present. This models the
colloidal suspension state on the substrate (regions with excess of
liquid). In this region the dynamics is governed by a hard-core
repulsion between spheres and a constant force from left to right
which is resulting from the liquid convection (liquid is flowing
from left to right). The pattern formation mechanism takes place in
region II, where the liquid level is lower than the height of the
nanospheres. This is the region that was considered in our earlier
model. Here, the dynamics is governed by capillary effects, friction
with the substrate and repulsion between disks, as
described earlier. When nanospheres get in this region, they 
become connected with springs having elastic constants $k_{ini}$. 
Finally, the third region (III.) represents the
final packed monolayer structure where no more forces are acting on
nanospheres. The separation line between region II. and III. defines
the drying front in the system.

\begin{figure}
\begin{center}
 \resizebox{8.5cm}{!}{
  \includegraphics{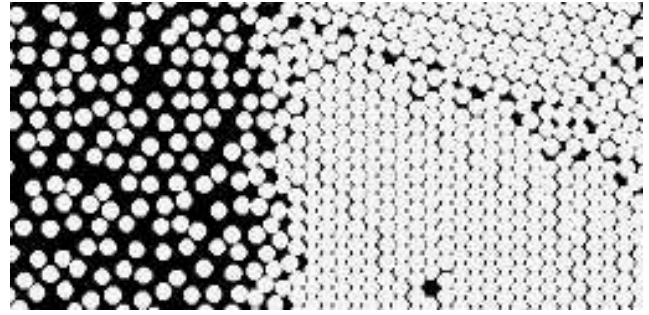}}
\end{center}
\caption{Snapshot from simulation. On the left side one can observe
the spheres in region I. and spheres in the right side of the
picture are already in region III. The drying front is also clearly
detectable.} 
\label{fig:6}
\end{figure}

\begin{figure}
\begin{center}
 \resizebox{7cm}{!}{
  \includegraphics{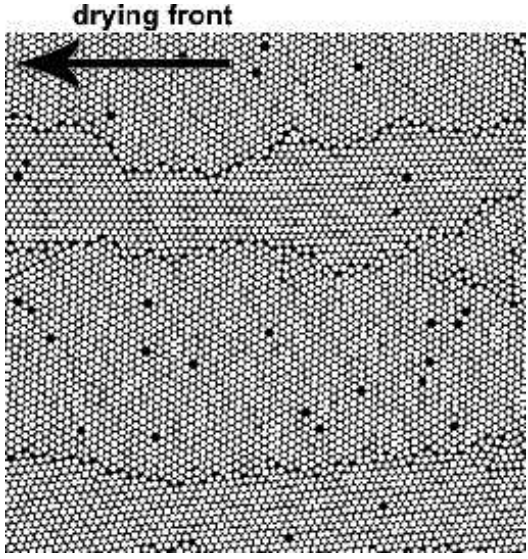}}
\end{center}
\caption{Typical structure obtained by computer simulations with the
improved (three region) model. Simulation parameters are: $F_f=0.003$,
$F_b=0.05$, $k_{ini}=0.01$, $dk=0.001$, the speed of the
interface between region I and region II was chosen to be
$10^{-6}*L/step$ ($L$ is the total length of the simulated region
and $steps$ stands for relaxation steps in the simulation). }
\label{fig:7}
\end{figure}

In the simulations initially we consider all nanospheres in region
I, and distribute them randomly with a somehow smaller density. We
define a given thickness for region II (usually 1/3 of the width of
the simulated region), and this region slowly progresses with a
constant speed from right to the left. Region I gets thus less and
less narrow and finally we end up with all spheres in region III.
During simulation new particles are constantly supplied from the
left and driven by the constant force in the direction of the
separation line between region I and region II. A snapshot from the
simulation can be observed in Figure \ref{fig:6} and a characteristic
final pattern is shown in Figure \ref{fig:7}. A qualitative
comparison between Figure \ref{fig:4} and Figure \ref{fig:7} will
convince us that this improved model works excellently and all type
of defects observable in experiments are reproduced by simulation.

\section{Effect of a random force}
\label{sec:3}

An extra shaking force acting on the nanospheres can be now imposed during the
simulated drying process. Two type of simulations were performed, using both the
original and the improved version of the spring-block model. In the original version
of the spring-block model we consider an extra stochastic force acting on each nanosphere.
This random force will model the influence of a random shaking applied on the drying
nanosphere suspension, and it approximates reasonably well the experimental conditions
of Sasaki and Hanne \cite{sasaki}, where an ultrasound bath was used to generate the shaking.
Second, we consider the improved model where in region II. an oscillating force is
applied on each nanosphere. This force is the same for all nanospheres and acts in
the direction of the drying front. This simulation corresponds to the
experimental setup of Sch\"ope \cite{schope}. In this experiment the inclined deposition method
was used for the deposition of the liquid film, and an extra oscillating
electric field perpendicular on the direction of the drag was applied on the nanospheres.

\subsection{The original model with an extra stochastic force}

\begin{figure}
\begin{center}
 \resizebox{8cm}{!}{
  \includegraphics{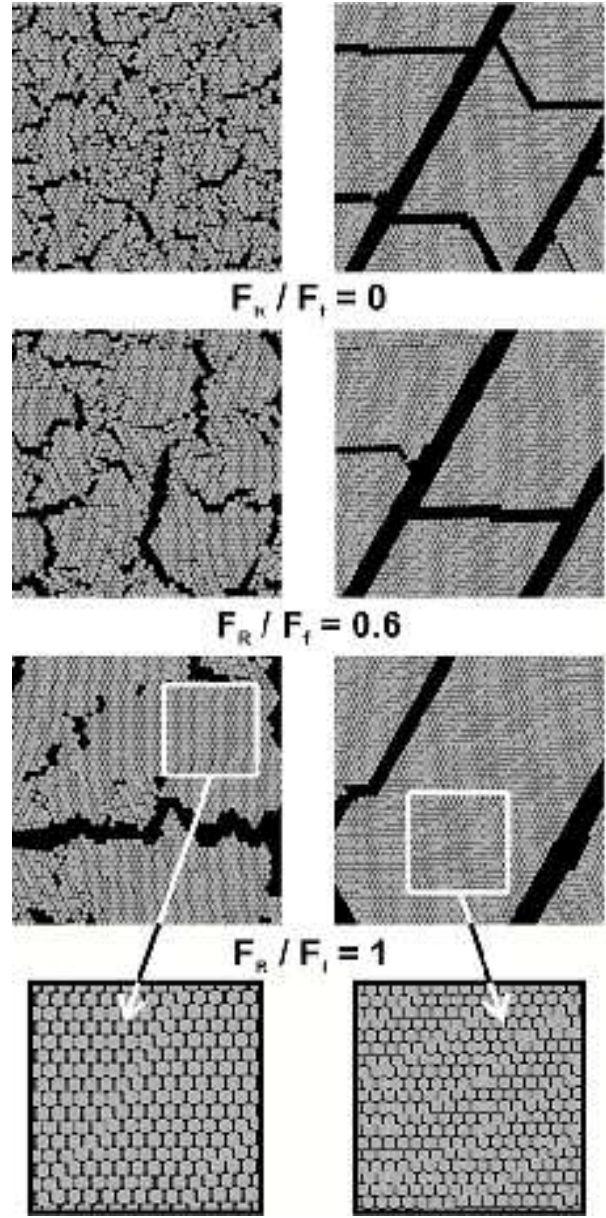}}
\end{center}
\caption{Influence of the strength of the applied random force on
the final stable patterns. Results for two different parameter sets
are shown. The pictures on the left side have been obtained with
$F_b = 0.05$ and the ones on the right side with $F_b = 0.1$. Other
parameters of both simulations are: $F_f=0.005$, $k_{\mathit{ini}} =
0.01$, $dk=0.001$ and $\rho = 0.749$.} 
\label{fig:8}
\end{figure}

An extra $F_r$ force completely uncorrelated in time and space was imposed on each nanosphere.
The orientation of this force was chosen to be totally random in the simulation plane
and it's strength was chosen with a uniform distribution in a fixed
$F_r \in [0, F_R]$ interval. The value of $F_R$ characterizes thus
the strength of the applied random perturbation. In order to achieve
a final stable configuration it is evident that the condition $F_R \le
F_f$ must be satisfied. Fixing the other parameters of the
simulations the influence of $F_R$ was systematically studied. Fixed
boundary conditions in a disk-like geometry was used and a central
rectangular region was studied. Extensive simulations proved that by
increasing the value of $F_R$ ($F_R \le F_f$) more and more ordered
structures are self-assembled. On Figure \ref{fig:8} we illustrate
this trend for two different parameter sets. Simulation results
confirm thus that a random shaking during the drying process could
be helpful in engineering more ordered structures. The effect is
reasonable, since the random force will reorder regions during the
whole simulation process. Dislocations, voids or fracture lines
formed in the early stages when many springs are present will
disappear during the dynamics, and the structure will evolve more
homogeneously. It is also observable from Figure \ref{fig:8} that
although larger ordered domains are generated with increasing $F_R$
values, the fracture lines becomes also wider. This effect is however
mainly due to the imposed fixed boundary conditions.

\subsection{The improved model with an extra oscillating force}

\begin{figure}
\begin{center}
 \resizebox{8.5cm}{!}{
  \includegraphics{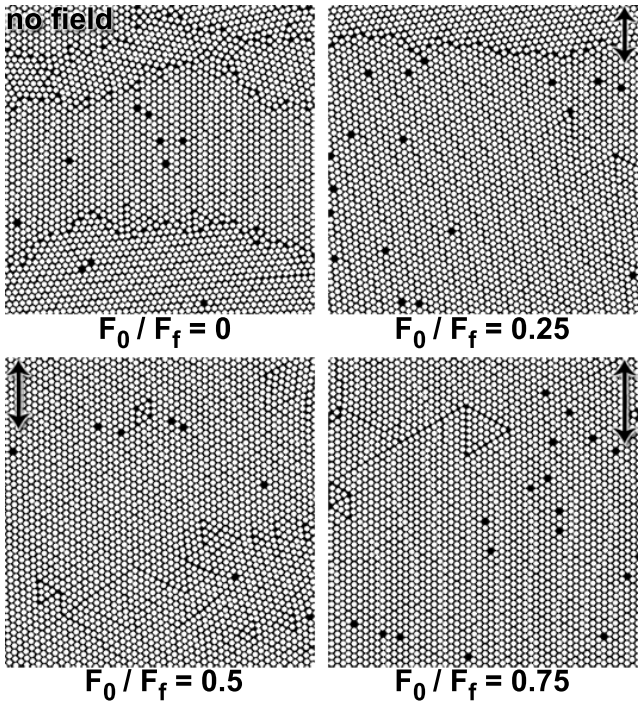}}
\end{center}
\caption{Final patterns obtained with an extra harmonic force in
region II. The successive pictures are for $F_0=0, F_f/4, F_f/2$ and
$3F_f/4$ amplitude values, respectively. The other simulation
parameters are as in Figure \ref{fig:8}. } 
\label{fig:9}
\end{figure}

Within the improved model we applied an additional harmonic $F_h=F_0
sin(\omega t)$ force in region II. The period of the harmonic force
was chosen to be of the order of hundreds of relaxation steps. The
$F_0$ amplitude of this force was systematically varied in the $F_0
\in [0, F_f)$ interval. The obtained results are similar with the
ones obtained with a random force. Increasing the value of $F_0$ in
the $[0, F_f)$  interval, the obtained structures become more
ordered and presents less defects. The result is stable and the
qualitative behavior does not change whenever the parameters of the
simulations are varied. A characteristic series of final patterns as
a function of $F_0$ is illustrated on Figure \ref{fig:9}. The
simulated series from Figure \ref{fig:9} describes well the
experimental results presented in \cite{schope}. Simulations
suggested also that higher frequencies eliminate more defects,
although the qualitative changes in the final structures are much
less evident than the results presented in Figure \ref{fig:9}. For a
systematic study on the frequency dependence simulations with much
bigger system sizes will be necessary.

\section{Experiments}
\label{sec:4}

As an experimental exercise we have also reproduced the results of 
Sasaki and Hane \cite{sasaki}.
Following their work the shaking was realized by an ultrasound radiation.
The experimental samples were prepared following the drop-coat
method \cite{Haynes2001,Kempa2003,Murray2004}. Of critical
importance for nano\-sphere ordering is the initial chemical
treatment of the glass substrate in order to render the surface
hydrophilic and improve its wettability (for details see
\cite{Jarai2005}). Polystyrene nanospheres of $400$
nm diameter, exhibiting negatively charged carboxyl-terminated
surface with a strongly hydrophobic nature, were supplied as
monodispersed suspensions in deionized water (wt 4\%). The original
suspension of polystyrene nanospheres was diluted by $10$ and a volume
of $100 \mu l$ diluted solution was evenly spread on the pre-treated
substrates. As the water evaporates and the samples are dried the
nanospheres self-assemble into close-packed monolayer arrays
exhibiting many fracture lines, dislocations or other type of
defects.

In order to investigate the influence of the shaking effect
drying was realized in two different conditions. Six samples were
studied, half of them were dried simply in the normal atmosphere of
the lab while the other half was exposed to a shaking effect induced
by an ultrasound bath.  

We have used an Elma Transsonic 35 kHz frequency ultrasound bath
filled with water. The glass plates holding the nanosphere solution
was positioned on the surface of the water by using a 10 cm diameter
and 0.4 cm thick cork floater with a disk-like geometry. In the
middle of the floater a hole with a 0.8 cm diameter was created,
allowing direct contact between the liquid from the ultrasonic bath
and the surface of the glass plates. Shaking is induced thus by the
acoustic vibrations transmitted through the glass plates.

In both cases (with and without ultrasound) the samples were
completely dried. Drying was achieved in approximately $45$ minutes
for the sonicated samples. For the nonsonicated samples complete
drying was achieved roughly two times slower. The microstructure of
the dried samples were studied by scanning electron microscopy (SEM)
using a JEOL JSM 5600 LV electronic microscope. By surveying
different regions of the substrates we have qualitatively analyzed
and compared the engineered nanosphere patterns.

\begin{figure}
\begin{center}
 \resizebox{8cm}{!}{
  \includegraphics{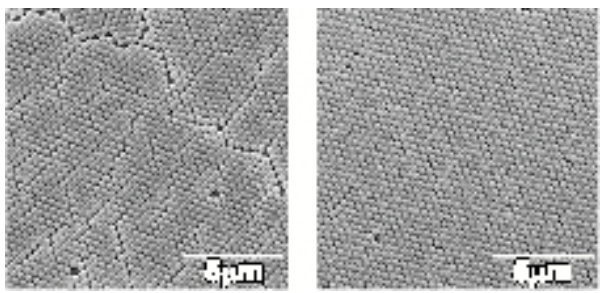}}
\end{center}
\caption{Visual comparison between characteristic regions of
non-sonicated (left) and sonicated samples (right).} 
\label{fig:10}
\end{figure}

Our experimental exercise confirms the picture suggested by Sasaki and Hane:
sonicated samples shows more extended and ordered triangular lattice structures
than non-sonicated ones. From a first
look on a randomly selected region on a sonicated and a
non-son\-i\-ca\-ted sample, this result is however not immediate.
One has to remember that due to the non-homogeneous spreading of the
liquid film, different regions in the same sample might exhibit
different level of ordering. In order to obtain thus a clear
conclusion many different regions have to be analyzed and a
statistical conclusion has to be drawn. We have analyzed tens of
different regions and qualitatively found that sonicated samples had
less number of fracture and dislocation line, the triangular lattice
structures being more extended in this case. For visually
illustrating this, on Figure \ref{fig:10} we present a
characteristic pattern found on sonicated samples in comparison with
a characteristic pattern found on non-sonicated samples.

\section{Conclusions}
\label{sec:5}

Nanospehere self-assembly from a drying suspension was studied in the presence 
of a moderate intensty shaking. To understand this pattern formation phenomenon 
two models were considered, (i) the simple Burridge-Knopoff 
type spring-block model \cite{Jarai2005} and (ii) an improved version of this model where the 
withdrawal of the liquid front and it's effect on self-assembly is also taken into account. 
In both cases beside the relevant capillarity, Coulombian and friction (pinning) forces 
and also an extra random shaking was imposed. Computer simulations suggested a clear conclusion: a moderate
intensity extra shaking imposed on the drying nanosphere system will yield
more ordered final patterns. Increasing the intensity of the random force up to the
limit of the pinning forces acting on the nanospheres is always beneficial and eliminates
most of the defects.  The simulation performed on our original spring-block model with an extra
random force describes well the experimental results of Sasaki and Hane \cite{sasaki}, while the
simulations realized on the improved model with an extra directional harmonic force acting in
region II corresponds to the experimental setup of Sch\"ope \cite{schope}. In agreement 
with experiments computer simulations also yield that primary (larger) fracture lines are formed   
predominantly in the direction of the moving liquid front (for experiments see Figure \ref{fig:4}
and for simulation Figure \ref{fig:6} or \ref{fig:7}). The patterns and the
trend obtained by computer simulations describes thus qualitatively well the experimentally
obtained ones, giving new evidences for the applicability of the simple
Burridge-Knopoff type models in modeling this system. 

We have not analyzed here the
influence of the frequency of the shaking. The first reason for this is that the influence
of the shaking frequency on the final pattern is less evident than the influence of
the intensity of the shaking force. In order to get reliable data computer simulations
on much larger systems have to be done. The computing power presently available for
us does not allow studying much larger systems in reasonable (months) computing time.
Second, the simulations in the present form lack real time. Instead of time we used the
relaxation steps for characterizing the dynamics. Relaxation steps might correspond however
to different time intervals and in this sense the frequency of the imposed oscillating force has only a qualitative meaning. 

As an exercise we have also repeated the experiments done by Sasaki and Hane,
confirming their results. During the experiments we realized that in order to get
more extended triangular lattice structures and less defects one will have to further 
optimize the experimental conditions. To make the experiments less costly computer
simulations could be useful. The present work shows that a simple Burridge-Knopoff type model 
works empirically well and can be helpful
to give a first estimate on the influence of many experimentally controllable parameters.

\section{Acknowledgments}
\label{ack}

The present work has been supported by the research grant CNCSIS
41/183 and CEEX-Nanobiospec 2006. The research of F. J\'arai-Szab\'o has also 
been supported by the Felowship Program for Transborder Hungarian Scientific Research - 
Hungarian Academy of Science. Discussions and helpful
comments from Prof. Peter Kralchevsky are gratefully acknowledged.

\end{document}